# Light Attenuation Length of High Quality Linear Alkyl Benzene as Liquid Scintillator Solvent for the JUNO Experiment

Hai-Bo Yang,[1] De-Wen Cao,[1] Chang-Wei Loh,[1] Ai-Zhong Huang,[2] Rui Zhang,[1] Yu-Zhen Yang,[1] Zhi-Qiang Qian,[1] You-Hang Liu,[1] Xue-Yuan Zhu,[1] Bin Xu,[1] Ming Qi,[1*]

1 National Laboratory of Solid State Microstructures and School of Physics, Nanjing University, Nanjing 210093, China
2 Jinling Petrochemical Corporation Ltd, Nanjing 210000, China

* Corresponding author: qming@nju.edu.cn

Abstract

The Jiangmen Underground Neutrino Observatory (JUNO) is a multipurpose neutrino experiment with a 20 kt liquid scintillator detector designed to determine the neutrino mass hierarchy, and measure the neutrino oscillation parameters. Linear alkyl benzene (LAB) will be used as the solvent for the liquid scintillation system in the central detector of JUNO. For this purpose, we have prepared LAB samples, and have measured their light attenuation lengths, with one achieving a length of 25.8 m, comparable to the diameter of the JUNO detector.

1. Introduction

The Jiangmen Underground Neutrino Observatory (JUNO) in Guangdong, China [1-3] is designed to consist of a spherically-shaped antineutrino detector, which is also the central detector, with a diameter of about 34 m, submerged in a water pool to reduce environmental radioactivity. The JUNO detector, which is under construction in Jiangmen, Guangdong, will measure the energy spectrum of the electron-antineutrinos coming from nuclear reactors at about 52 km from the detector to determine the neutrino mass hierarchy, and perform sub-percent precision measurements on the value of the neutrino-mixing parameters [4]. By resolving the mass hierarchy, in particular the relative masses of the only three neutrinos currently known in the Standard Model, we may gain a better understanding of the nature of the neutrino flavor conversions and density of isotopes in supernova explosions [5, 6].

To achieve the physics purposes of JUNO, about 20 kilotons of linear alkyl benzene (LAB) with a formula of $C_6H_5C_nH_{2n+1}$ (n=10~13) [7] is projected to be filled into the interior of the antineutrino detector as the detection medium, and as a solvent for a liquid scintillation system comprising of 2,5-diphenyloxazole as the fluor, and p-bis-(o-methylstyryl)-benzene as the wavelength shifter. The detector will be installed with 17k photomultiplier tubes (PMT) to detect the emitted photons from the interaction between the positrons and neutrons produced from inverse beta decays and the liquid scintillator. For the emitted photons to reach the PMTs from their production vertex located within the detector, good optical transparency is needed from the LAB. For this purpose, the transparency of the LAB is quantified by its light attenuation length. The LAB samples for use in the JUNO experiment should achieve at least an attenuation length of 30 m, which is about the diameter of the spherical detector. As a comparison, the Daya Bay experiment [8-12] with a diameter of 5 m and a height of 5 m cylindrical detector contains LAB with an attenuation length of about 10 m.

In this work, numerous LAB samples have been prepared by the Jinling Petrochemical Corporation (hereafter known as "Jinling"), some potentially scalable for a mass production of 20 kilotons of LAB for the purpose of the JUNO experiment. We have successfully measured such samples up to an attenuation length of 25 m, nearing the target length of the JUNO experiment demand. A preliminary study has also been performed on the organic impurity content of the samples used in this work, with some of these impurities have been found to be able to reduce the optical transparency of the LAB.

## 2. LAB Samples Preparation

LAB is the most common raw material in the manufacturing of biodegradable household and industrial detergents, produced by the alkylation of benzene starting from a kerosene feedstock. LAB is proposed as a liquid scintillator solvent in JUNO for its attractive properties, including rich H atoms content, appreciable light yield, high flash point (130°C, which can significantly reduce safety concerns for the JUNO experiment), relatively inexpensive, non-toxic and hence environmentally friendly. Apart from JUNO and Daya Bay, LAB has also been widely used in past antineutrino experiments including KamLAND [13], Borexino [14] and RENO [15]. The most important experimental requirement for a large scale usage of LAB in the JUNO experiment is a high light attenuation length within a wavelength region of 350 – 550 nm. Unfortunately, such high quality LAB is not available from the commercial-based production of LAB.

Some samples of high quality LAB were prepared with different methods, and tagged as NJ32#, NJ33#, NJ41#, NJ42#, NJ43# and NJ44#. The NJ32# sample is a highly-purified sample prepared with brute force aluminum oxide filtration technique at the manufacturing site of Jinling, whereas NJ33#, NJ41#, NJ#42, NJ43# and NJ44# were prepared with an improvised LAB production process without the brute force filtration method, which is more suitable for large-scale mass production in the future. Among them, NJ44# sample is the better result of this latest improvised technique yet.

## 3. Experiment

### 3.1. Attenuation Length of LAB

For an initial intensity $I_0$ of an incident light at position x = 0 propagating with a wavelength $\lambda$ in a liquid material comprising of a mixture of N attenuating species, with the intensity reduced to $I$ after propagating an x distance through the liquid, the expression for the intensity can be written as follows [16]:

$$I = I_0 e^{-\frac{x}{L_\lambda}}, \qquad (1)$$

where the attenuation length $L_\lambda$ in Equation (1) is related to the attenuation cross section of species i, $\sigma_i$, and the number density of species i, $n_i$ as follows:

$$L_\lambda = \frac{1}{\sum_{i=1}^{N} \sigma_i n_i} \quad . \qquad (2)$$

Even for a highly-purified LAB sample, there are still impurities at the order of several ppb and ppt, and hence the attenuation length of the LAB samples will still be influenced by the impurities.

### 3.2. Experimental Setup

The experimental setup is composed of two parts: the optical measuring and controlling system, and the data acquisition system (DAQ) [17]. Figure 1 and 2 shows the optical and DAQ system of the experiment respectively.

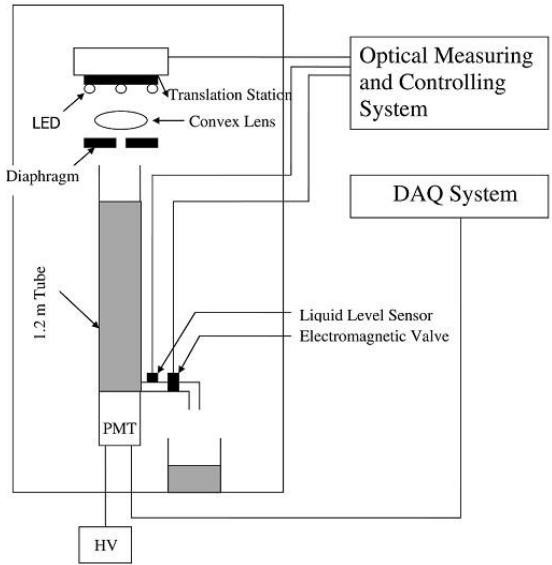

Figure 1: Diagram of the experimental setup.

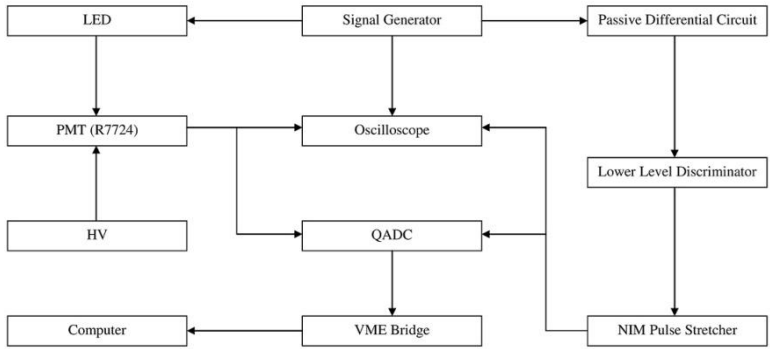

Figure 2: Schematic diagram of the DAQ system.

Light pulses with a 430 nm mean wavelength from an LED source were propagated through a convex lens and a diaphragm with an adjustable aperture system to aim towards the LAB liquid contained in a vertically-erected 1.2 m tube with Teflon-covered inner wall. A PMT was placed at the bottom of the tube to capture the attenuated outgoing light. The entire optical measuring and controlling system was placed in a dark room to reduce stray light impinging the PMT. The height of the LAB liquid was controlled by an electromagnetic valve at the bottom of the tube. The pulse signal of the collected light from the PMT was transmitted to the DAQ system via a charge-to-amplitude converter, QADC (CAEN V965), to be recorded as an ADC channel value proportional to the intensity of the collected light. In this manner, Equation (1) can be rewritten as:

$$ADC_x = ADC_0 e^{-\frac{x}{L_\lambda}}, \qquad (2)$$

where $ADC_x$ is the ADC value when the height of the LAB is $x$, and $ADC_0$ is the ADC value when the tube is empty of any liquid.

3. 3. Optical System Calibration

The light beam was calibrated to focus towards the cylindrical tube via the small aperture of the diaphragm, and thereafter towards the central area of the PMT. This was done by calibrating the relative position and distance between the light source and the lens system. The aperture was adjusted to a diameter of 1 mm, the maximum size possible to avoid PMT saturation.

3. 4. Parameter Tuning of the Experiment

Purified water with a high attenuation length was used initially in the experiment for calibration purposes, with the optical and DAQ system obtaining optimal light collection and signal output while simultaneously ensuring that the resulting output was still within the accepted range of the V965 converter. This was done, among others, by controlling the incoming light intensity through the voltage given to the LED, and the adjustment for the aperture. To optimize the signal-to-noise ratio, the driving pulse rate of the LED was controlled through adjusting the duty ratio of the waveform produced from the signal generator. The ambient temperature of the experiment was stabilized to within 0.2 – 0.5 % during the attenuation length measurements of the samples to reduce fluctuations in the output ADC channel values. Programs written in LabVIEW were used to control the readout of the DAQ system, and for analysis of the data.

3. 5. PMT Selection

A PMT model CR135 was initially used for NJ32# and NJ33#, and a PMT model R7724 was later used for the other LAB samples, where the latter had a faster signal response than the former. Figure 3 shows a scope view of a signal from the PMT model CR135 and R7724 in the data acquisition window. It can be seen that the after-pulses and noise are not significant in the latter compared to the former.

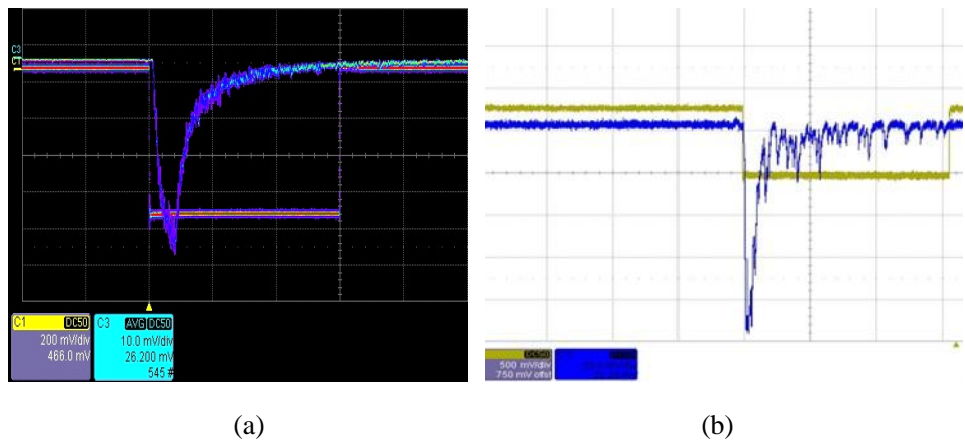

(a) (b)

Figure 3 (a),(b): A scope view of the signal from the PMT model R7724 and CR135 respectively.

4. Results and Discussion

4. 1. Attenuation Length Measurement

Figure 4 shows the attenuation length as obtained from the exponential fit to the data for NJ32#, NJ33# and NJ44#. The exponential fit for NJ44# was in better agreement with the data than those for NJ32# and NJ33#. Compared to NJ32# and NJ33#, the attenuation length experiment of NJ44# was performed with the new PMT R7724 model replacing the PMT CR135 model. The ambient temperature was controlled to a smaller fluctuation range for NJ44# compared to the other two samples.

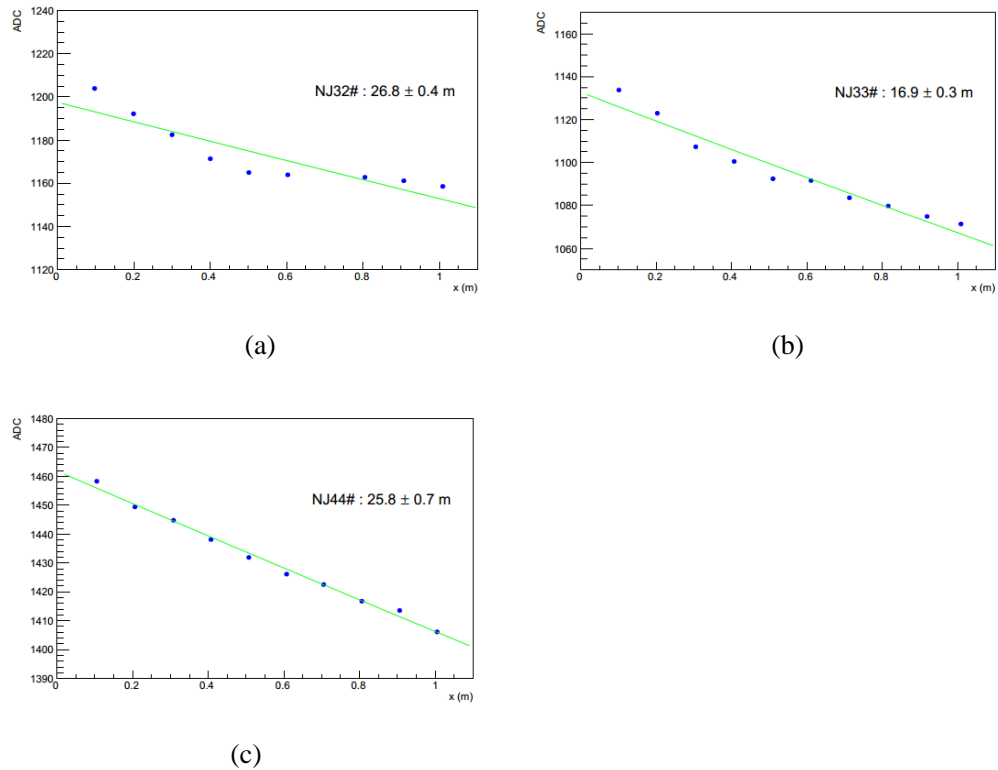

(a)

(b)

(c)

Figure 4 (a), (b) and (c): Results for NJ32#, NJ33# and NJ44# respectively for various LAB liquid height x. The attenuation lengths, as obtained from the exponential fit, are 26.8 m, 16.9 m and 25.8 m respectively.

4. 2. Improvement in Preparation Process

Figure 5 shows the measurement results for NJ41#, NJ42#, NJ43# and NJ44# and their corresponding fit to the data. As these samples were prepared with the same method but with different preparation parameters, we have also studied the correlation between their acid-washed colorimetry values [18] and their corresponding attenuation lengths. In the acid-washed colorimetry method, sulfuric acid is mixed into the LAB, and is then left to stand for five minutes. A filtration is then performed on the resulting mixture. The light absorption strength of the LAB relative to that of a distilled water sample with high purity is computed as a wavelength-specific value. Figure 6 shows a graph of a trend in the attenuation length with respect to the acid-washed colorimetry value of NJ41# to NJ44#. In this work, we find that the acid-washed colorimetry value correlates positively with the attenuation length of these LAB samples for a 430 nm wavelength light. This forms a guideline for future preparation of LAB samples to achieve the JUNO experiment aim of an attenuation length of at least 30 m as the acid-washed colorimetry technique can be performed in-situ during the production of the LAB. Furthermore, the results of NJ41# to NJ44# convinced us that the tuning of the production parameters is in the right direction as the attenuation lengths, as computed from the fits to the data obtained using the aforementioned experimental setup, are increasing.

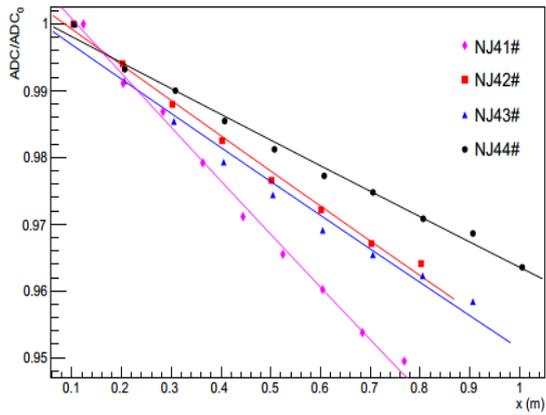

Figure 5: ADC/ADC$_0$ of NJ41#, NJ42#, NJ43# and NJ44# for various LAB liquid height x, where ADC/ADC$_0$ refers to the ADC value at height x normalized to the ADC value at height x$_0$ which is the largest value as obtained from the data.

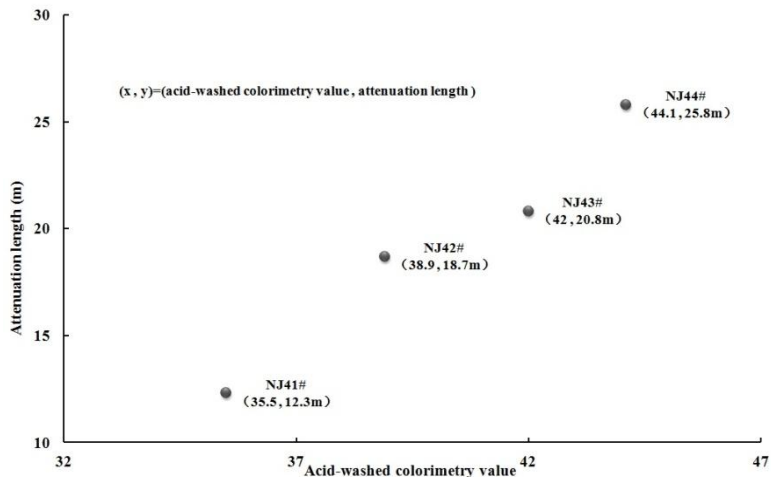

Figure 6: Correlation between the attenuation length and the acid-washed colorimetry (AWC) value for various LAB samples, with their values indicated as (AWC value, attenuation length) in the Figure.

4. 3. Organic Impurity Analysis

To achieve an attenuation length of more than 25 m, it is imperative that the organic impurities in the LAB samples are identified, in particular those that have a light absorption tendency at a visible wavelength of about 430 nm which is the envisaged working point wavelength of the PMTs to be used in the JUNO experiment. In our previous works, we have investigated the possibilities of inorganic [19] and organic [20, 21] impurities on the absorption of visible wavelength light at about 430 nm and their consequences to the attenuation length our LAB samples. In samples NJ32#, NJ33# and NJ44#, we have used gas chromatography technique to investigate their impurities through their light absorption spectrum, and found that they contain several impurities in the order of several ppb in concentrations, in particular one with a molecular formula of $C_{34}H_{36}N_2O_4$ with a strong absorption at a visible wavelength of about 420 nm. Investigations are currently being conducted on the organic impurities in the LAB samples to understand their structures and their influence on the overall optical transparency of the samples. This would further help us in modifying the LAB preparation technique for mass production thereof, and thus improving the transparency of the high quality LAB samples in the near future.

4.4. Further Improvement to the Experiment

To further accurately determine the attenuation lengths of the samples, we have identified several aspects of the experiment that will need upgrades in the future. As Equation (1) is ideally used with a single wavelength incoming light and hence the LED used in our optical system is not as desirable, it will be substituted in future experiments with a monochromatic light source, for example a suitable laser diode. In the current DAQ system, we have used a V965 model, a charge-to-amplitude converter with one gate input, as can be observed from Figure 3, which we had adjusted its activation to correspond to the PMT signal timing. Our plan in future is to update the converter with a waveform digitizer, such as a module CAEN 1729 or a Flash ADC, which would among others, assist in reducing the amount of noise by determining in a better manner when the DAQ is receiving an output above its pedestal baseline.

## 5. Summary

In this work, we have measured the attenuation length of our LAB samples, with one of them obtaining 25.8 m with our calibrated experimental setup and PMT model R7724. The highest attenuation length has been obtained from a sample prepared with a method suitable for large-scale mass production. A positive correlation has been found between the acid-washed colorimetry value and the attenuation length at 430 nm light wavelength of the LAB samples. This result is useful as a guideline to prepare future LAB samples with higher attenuation length and excellent optical properties in order to meet the target of the JUNO experiment. Further investigation is currently being done on the organic impurities in the LAB samples, and their light absorption properties particularly at the wavelength of 430 nm. Although these impurities are in the order of several ppb and some even of several ppt in concentration in the LAB samples, they may be the key in achieving our target of at least 30 m in attenuation length.

**Disclosure**

Haibo Yang and Dewen Cao are considered co-first author.

**Acknowledgments**

We are grateful for warm help and enlightening insights from Wang Yi-Fang, Cao Jun, Qian Sen, Zhou Li, Ding Ya-yun, Ning Zhe, Zhu Na, Yu Guang-You and Wang Wei. This work was supported by the National 973 Project Foundation of the Ministry of Science and Technology of China (Contract No. 2013CB834300).

**Competing Interests**

The authors declare that they there are no competing interests regarding the publication of this paper.

**References**

1. L. Zhan, Y. F. Wang, J. Cao et al, "Determination of the neutrino mass hierarchy at an intermediate baseline," *Physical Review*, vol. 78, Article ID 111103(R), 2008.

2. L. Zhan, Y. F. Wang, J. Cao et al, "Experimental requirements to determine the neutrino mass hierarchy using reactor neutrinos," *Physical Review*, vol. 79, Article ID 073007, 2009.

3. Y. F. Li, J. Cao, Y. F. Wang et al, "Unambiguous determination of the neutrino mass hierarchy using reactor neutrinos," *Physical Review*, vol. 88, Article ID 013008, 2013.

4. F. P. An, G. P. An, Q. An et al, "Neutrino physics with JUNO," *Journal of Physics G: Nuclear and Particle Physics*, vol. 43, No. 3, Article ID 030401, 2016.

5. K. C. Lai, F. F. Lee, F. S. Lee et al, "Probing Neutrino Mass Hierarchy by Comparing the

Charged-Current and Neutral-Current Interaction Rates of Supernova Neutrinos," *Journal of Cosmology and Astroparticle Physics*, vol. 07, Article ID 039, 2016.

6. T. Kajino, W. Aoki, M. K. Cheoun et al, "Supernova constraints on neutrino oscillation and EoS for proto-neutron star," *AIP Conference Proceedings*, vol. 1594, pp. 319, 2014.

7. W. Konetschny and W. Kummer, "Nonconservation of total lepton number with scalar bosons," *Physics Letters B*, vol. 70, pp. 433-435, 1977.

8. F. P. An et al (Daya Bay Collaboration), "Spectral Measurement of Electron Antineutrino Oscillation Amplitude and Frequency at Daya Bay," *Physical Review Letters*, vol. 112, Article ID 061801, 2014.

9. F. P. An et al (Daya Bay Collaboration), "Search for a Light Sterile Neutrino at Daya Bay," *Physical Review Letters*, vol. 113, Article ID 141802, 2014.

10. F. P. An et al (Daya Bay Collaboration), "New Measurement of Antineutrino Oscillation with the Full Detector Configuration at Daya Bay," *Physical Review Letters*, vol. 115, Article ID 111802, 2015.

11. F. P. An et al (Daya Bay Collaboration), "Measurement of the Reactor Antineutrino Flux and Spectrum at Daya Bay," *Physical Review Letters*, vol. 116, Article ID 061801, 2016.

12. F. P. An et al (Daya Bay Collaboration), "Improved Search for a Light Sterile Neutrino with the Full Configuration of the Daya Bay Experiment," *Physical Review Letters*, vol. 117, Article ID 151802, 2016.

13. S. Abe et al (The KamLAND Collaboration), "Precision Measurement of Neutrino Oscillation Parameters with KamLAND," *Physical Review Letters*, vol. 100, Article ID 221803, 2008.

14. C. Arpesella et al (Borexino Collaboration), "Direct Measurement of the 7Be Solar Neutrino Flux with 192 Days of Borexino Data," *Physical Review Letters*, vol. 101, Article ID 091302, 2008.

15. J. K. Ahn et al (RENO Collaboration), "Observation of Reactor Electron Antineutrinos Disappearance in the RENO Experiment," *Physical Review Letters*, vol. 108, Article ID 191802, 2012.

16. F. Baldini, A. Giannetti, "Optical chemical and biochemical sensors: new trends," *Optical Sensing and Spectroscopy*, vol. 5826, pp. 485, 2005.

17. Z. Ning, S. Qian, Z. W. Fu et al "A data acquisition system based on general VME system in WinXP," *Nuclear Techniques,* vol. 33(10), pp. 740–744, 2010. (in Chinese)

18. A. Z. Huang, P. C. Hu, Z. W. Fu et al "Application of the acid-washed colorimetry measurement on linear alkyl benzene (lab) quality control," *Chemical Analysis & Meterage*, vol. 20, No. 6, pp. 43-47, 2011. (in Chinese)

19. G. Y. Yu, D. W. Cao, A. Z. Huang et al "Some new progress on the light absorption properties of linear alkyl benzene solvent ," *Chinese Physics C* , vol. 40, No. 1, Article ID 016002, 2016.

20. P. W. Huang, P. Y. Li, Z. W. Fu et al "Study of attenuation length of linear alkyl benzene as LS solvent ," *Journal of Instrumentation*, vol. 5, Article ID P08007, 2010.

21. P. W. Huang, H. Y. Cao, M. Qi et al, "Theoretical study of UV-Vis light absorption of some impurities in alkyl benzene type liquid scintillator solvent," *Theoretical Chemistry Accounts*, vol. 129, pp. 229–234, 2011.